\documentstyle[11pt,aaspp4]{article}
\begin{document}
\def\hh{\, h^{-1}}
\newcommand{\wth}{$w(\theta)$}
\newcommand{\xir}{$\xi(r)$}
\newcommand{\Lya}{Ly$\alpha$}
\newcommand{\Lyb}{Lyman~$\beta$}
\newcommand{\Hb}{H$\beta$}
\newcommand{\HI}{H{\sc I}}
\newcommand{\msun}{M$_{\odot}$}
\newcommand{\sfr}{M$_{\odot}$ yr$^{-1}$}
\newcommand{\sfrd}{M$_{\odot}$ yr$^{-1}$ Mpc$^{-3}$}
\newcommand{\cld}{erg s$^{-1}$ Hz$^{-1}$ Mpc$^{-3}$}
\newcommand{\dnsty}{$h^{-3}$Mpc$^3$}
\newcommand{\za}{$z_{\rm abs}$}
\newcommand{\ze}{$z_{\rm em}$}
\newcommand{\cmtwo}{cm$^{-2}$}
\newcommand{\nhi}{$N$(H$^0$)}
\newcommand{\degpoint}{\mbox{$^\circ\mskip-7.0mu.\,$}}
\newcommand{\halpha}{\mbox{H$\alpha$}}
\newcommand{\hbeta}{\mbox{H$\beta$}}
\newcommand{\hgamma}{\mbox{H$\gamma$}}
\newcommand{\kms}{\,km~s$^{-1}$}      
\newcommand{\minpoint}{\mbox{$'\mskip-4.7mu.\mskip0.8mu$}}
\newcommand{\mv}{\mbox{$m_{_V}$}}
\newcommand{\Mv}{\mbox{$M_{_V}$}}
\newcommand{\peryr}{\mbox{$\>\rm yr^{-1}$}}
\newcommand{\secpoint}{\mbox{$''\mskip-7.6mu.\,$}}
\newcommand{\sqdeg}{\mbox{${\rm deg}^2$}}
\newcommand{\squig}{\sim\!\!}
\newcommand{\subsun}{\mbox{$_{\twelvesy\odot}$}}
\newcommand{\et}{{\it et al.}~}
\newcommand{\er}[2]{$_{-#1}^{+#2}$}
\def\h50{\, h_{50}^{-1}}
\def\hbl{km~s$^{-1}$~Mpc$^{-1}$}
\def\ltsima{$\; \buildrel < \over \sim \;$}
\def\simlt{\lower.5ex\hbox{\ltsima}}
\def\gtsima{$\; \buildrel > \over \sim \;$}
\def\simgt{\lower.5ex\hbox{\gtsima}}
\def\arcs{$''~$}
\def\arcm{$'~$}
\newcommand{\wu}{$U_{300}$}
\newcommand{\wb}{$B_{435}$}
\newcommand{\wv}{$V_{606}$}
\newcommand{\wi}{$i_{775}$}
\newcommand{\wz}{$z_{850}$}
\newcommand{\hmpc}{$h^{-1}$Mpc}
\newcommand{\um}{$\mu$m}
\title{The Rest--Frame UV Luminosity Density of Star--Forming Galaxies at 
 Redshifts $z>3.5$\altaffilmark{1}} 
\author{\sc M. Giavalisco\altaffilmark{2}, M. Dickinson\altaffilmark{2,3}, 
  H. C. Ferguson\altaffilmark{2,3}, S. Ravindranath\altaffilmark{2}, 
  C. Kretchmer\altaffilmark{3}, L. A. Moustakas\altaffilmark{2}, 
  P. Madau\altaffilmark{4}, S. M. Fall\altaffilmark{2}, 
  Jonathan P. Gardner\altaffilmark{5}, M. Livio\altaffilmark{2}, 
  C. Papovich\altaffilmark{6}, A. Renzini\altaffilmark{7}, 
  H. Spinrad\altaffilmark{8}, D. Stern\altaffilmark{9}, 
  A. Riess\altaffilmark{2}} 
\affil{$^{2}$Space Telescope Science Institute, 3700 San Martin Dr.,
       Baltimore, MD 21218} 
\affil{$^{3}$Department of Physics and Astronomy, The Johns Hopkins
       University, 3400 N. Charles St., Baltimore, MD 21218--2686}
\affil{$^{4}$Department of Astronomy and Astrophysics, University of
       California Santa Cruz, 1156 High St., Santa Cruz, CA 95064} 
\affil{$^{5}$Goddard Space Flight Center, Code 681, Greenbelt, MD 20771}
\affil{$^{6}$Steward Observatory, University of Arizona, 933 Cherry Ave.,
       Tucson, AZ 85721--0065}
\affil{$^{7}$European Southern Observatory, Karl Schwarzschild Strasse 2,
       85748, Garching, Germany}
\affil{$^{8}$Department of Astronomy, University of California Berkeley,
       Berkeley, CA 94720}
\affil{$^{9}$Jet Propulsion Laboratory, California Institute of Technology,
       Mail Stop 169-327, Pasadena, CA 91109}

\altaffiltext{1}{Based on observations obtained with the NASA/ESA {\it Hubble
Space Telescope} obtained at the Space Telescope Science Institute, which is
operated by the Association of Universities for Research in Astronomy, Inc. 
(AURA) under NASA contract NAS 5-26555.}. 

\begin{abstract}

We have measured the rest--frame $\lambda\sim 1500$ \AA\ comoving specific
luminosity density of star--forming galaxies at redshift $3.5<z<6.5$ from deep
images taken with the {\it Hubble Space Telescope} ({\it HST}) and the
Advanced Camera for Surveys (ACS), obtained as part of the Great Observatories
Origins Deep Survey (GOODS). We used color selection criteria to construct
samples of star--forming galaxies at redshifts $z\sim 4$, $5$ and $6$,
identified by the signature of the 912 \AA\ Lyman continuum discontinuity and
\Lya\ forest blanketing in their rest--frame UV colors (Lyman--break
galaxies). The ACS samples cover $\sim 0.09$ square degree, and are also
relatively deep, reaching between 0.2 and 0.5 $L_3^*$, depending on the
redshift, where $L_3^*$ is the characteristic UV luminosity of Lyman--break
galaxies at $z\sim 3$. The specific luminosity density of Lyman--break
galaxies appears to be nearly constant with redshift from $\approx 3$ to
$z\sim 6$, although the measure at $z\sim 6$ remains relatively uncertain,
because it depends on the accurate estimate of the faint counts of the $z\sim
6$ sample.  If Lyman--break galaxies are fair tracers of the cosmic star
formation activity, our results suggest that at $z\sim 6$ the universe was
already producing stars as vigorously as it did near its maximum several Gyr
later, at $1\simlt z\simlt 3$. Thus, the onset of large--scale star formation
in the universe is to be sought at around $z\sim 6$ or higher, namely at less
than $\sim 7$\% of the current cosmic age.
\end{abstract}
\keywords{cosmology: observations --- galaxies: formation --- galaxies: 
evolution --- galaxies: distances and redshifts}

\section{INTRODUCTION}

The amount of star formation that took place early in the cosmic history,
e.g. at $z>2$, and its evolution with redshift are key pieces of information
to constrain theories of galaxy evolution. Evidence of the possible signature
of the trailing edge of the cosmic reionization epoch at redshifts $z\simgt 6$
(e.g. Fan et al. 2002) has also recently renewed interest in measuring the star
formation activity at these epochs, since starburst galaxies can contribute a
significant fraction of the ionizing photons (Madau et al. 1999; Haiman et
al. 2001; Steidel et al. 2001). Initial estimates based on the evolution of
the cosmic neutral hydrogen mass density estimated from damped \Lya\
absorption systems (DLA, Fall et al. 1996) suggested that star formation
activity peaked in the past at around $z\sim 1$ and then decreased at higher
redshifts. The first direct measures based on the observed UV radiation of
relatively unobscured star--forming galaxies in the local and
intermediate--redshift universe and of Lyman--break galaxies (LBGs) at $z>2$
(Madau et al. 1996) seemed to confirm this general scenario. However,
subsequent measures based on the systematic spectroscopic identification of
hundreds of Lyman--break galaxies at $z\sim 3$ and $z\sim 4$ showed that the
star formation density traced by these sources, after reaching a maximum value
somewhere between $z\sim 1$ and $\sim 2$, remains nearly constant up to the
highest redshift observed with some confidence, i.e. $z\sim 4$ (Steidel et
al. 1999, S99 hereafter). More recent measure at $z\sim 5$ from deep,
wide--area samples obtained with the Subaru telescope (Iwata et al. 2003) and
with {\it HST} (Bouwens et al. 2003) show very mild evolution from $z\sim 3$,
although possible evidence to the contrary has also been reported (Stanway et
al. 2003a,b). Measures based on photometric redshifts also seem to be in
general agreement with a constant or mildly decreasing activity of star
formation as traced by starburst galaxies up to $z\sim 6$ (Thompson et
al. 2001; Thompson 2003; Fontana et al. 2003; Kashikawa et al. 2003), or even
suggest that the star--formation activity might actually increase with
redshift (Lanzetta et al. 2002).

The current measures of UV specific luminosity density of LBGs come from the
Hubble Deep Field survey (HDF, Williams et al. 1996, 2000) and other deep, but
relatively small survey made with {\it HST} (Bouwens et al. 2003) or from
ground--based surveys. The {\it HST} surveys are limited by its very small
area, and thus cosmic volume, which makes it prone to systematic bias induced
by cosmic variance (for example, S99 argue that the original result by Madau
et al. (1996) was, at least in part, the effect of cosmic fluctuations). The
ground--based surveys cover a much larger area, but at high redshifts the
brightness of the night sky limits the sensitivity to only the bright end of
the luminosity function. Unfortunately, LBGs are rather faint, with a
characteristic magnitude $z_{850}^*=26.0$ at $z=6$ (based on Adelberger \&
Steidel 2000's UV luminosity function at $z=3$) if no luminosity evolution
occurs between $z\sim 3$ and $z\sim 6$. Thus, one needs to reach rather faint
flux limits, even for 8--m telescopes, to account for most of the light from
LBGs once the redshift exceeds $z\sim 4$.

The GOODS/ACS observations offer a good compromise, covering an area $\sim 33$
times larger than the combined HDFs, with much deeper sensitivity out to
wavelengths of nearly 1 $\mu$m than that of most ground--based samples. This
makes it possible to detect galaxies with luminosity fainter than $L_3^*$ up
to $z\sim 6.5$. In this letter we present measures of the comoving specific
luminosity density over the range $3.5\simlt z\simlt 6.5$ based on the stack
of three epochs of observations in both GOODS fields, the HDF--N and CDF--S. A
companion letter by Dickinson et al. (2003, D03 hereafter) in this volume
analyzes in greater detail the selection of galaxies at $z\sim 6$, including a
discussion of the first spectroscopic identifications. Throughout, magnitudes
are in the AB scale (Oke 1977), and the world model, when needed, is that of a
flat universe with density parameters $\Omega_m=0.3$, $\Omega_{\Lambda}=0.7$
and Hubble constant $H_0=70$ \hbl.

\section{DATA AND SAMPLE SELECTION}

The data used in this letter are comprised of the first three epochs of
observations, out of a total of five, of the GOODS/ACS survey (Giavalisco et
al. 2003). They consist of eight mosaics, four in each of the two survey
fields, and one in each of the \wb, \wv, \wi\ and \wz\ bands. Each mosaic
covers an area approximately 10'x16' in size with a scale of 0.05
arcsec/pixel. Table 1 of Giavalisco et al. (2003) lists the relevant
parameters of the data, including the sensitivity. Source catalogs have been
extracted using the SExtractor software, and the procedures are also described
in Giavalisco et al. (2003).  The catalogs used to select high--redshift
galaxies have been created performing the source detection in the \wz\ band
and then using the isophotes defined during this process as fixed apertures
for photometry in the other bands. We derive colors from isophotal magnitudes,
but use the AUTO aperture--corrected magnitudes (in the SExtractor parlance)
to measure the \wz--band apparent magnitudes.

Samples of star--forming galaxies at high redshifts have been extracted from
the matched catalogs using the Lyman--break technique (Madau et al. 1996;
Dickinson 1998; S99, Steidel et al. 2003; see also Giavalisco 2002). 
Specifically, we have defined LBGs at $z\sim 4$ (\wb--band dropouts) using the
colors equations:
$$(B_{450}-V_{606})\ge 1.2 + 1.4\times (V_{606}-z_{850})~~~\wedge~~~ 
(B_{450}-V_{606})\ge 1.2~~~\wedge~~~ (V_{606}-z_{850})\le 1.2\eqno(1.1),$$
and LBGs at $z\sim 4$ (\wv--band dropouts) by: 
$$[(V_{606}-i_{775})>1.5+0.9\times (i_{775}-z_{850})]~~~\vee$$
$$\vee~~~[(V_{606}-i_{775})>2.0]~~~\wedge~~~(V_{606}-i_{775})\ge 1.2~~~
\wedge~~~(i_{775}-z_{850})\le 1.3\eqno(1.2),$$
where $\vee$ and $\wedge$ are the logical OR and AND operators, respectively.
For bands which are entirely shortward of the Lyman discontinuity (e.g. the
\wb\ for the \wv-- and \wi--band dropouts and the \wv\ band for the \wi--band
dropouts) we also request a {\it non detection} with $S/N<2$. 

These selection criteria are largely based on our previous experience with
Lyman--break galaxies (e.g. Dickinson 1998; S99; Steidel et al. 2003). We have
visually fine--tuned them based on the observed colors of stars and galaxies
in the ACS images, as well as on the ACS synthetic photometry from galaxy
spectral templates to reject most interlopers from lower redshifts, while
efficiently detecting typical UV--bright, star--forming galaxies at redshifts
of interest.

Without a third band in the near--infrared, it is not possible to use a 
2--color selection to define samples of LBGs at $z\sim 6$ in analogy to what
done above. Instead, we have used the single--color threshold 
$$(i-z)\ge 1.3,\eqno(1.3)$$ 
along with a non--detection requirement $S/N<2$ in the \wb\ and \wv\ bands to
define our sample of $z\sim 6$ \wi--band dropouts. D03 discuss the selection
and robustness of these dropouts in detail, including the use of available
near--IR photometry as ``a posteriori test'' of the selection criteria. A
number of galaxies meeting these criteria have already been confirmed
spectroscopically to be at $z\sim 6$ (Bunker et al. 2003; D03; Stanway et
al. 2003b). Later on we discuss the implications of this different selection
of LBGs on the measure of the specific luminosity density of the \wi--band
dropouts.

We have included in the samples only galaxies with $S/N\ge 5$ in the \wz\
band, and we have visually inspected each of them and removed sources that
were deemed as not real, either artifacts or spurious detections (estimated
using counts of negative sources detected in the same data set). These amount
to a negligible number for the \wb\ and \wv\ samples, and $\approx 12$\% for
the \wi\ one. We have also eliminated from all samples sources with stellar
morphology down to apparent magnitude \wz$\sim 26$, i.e. where such a
morphological classification is reliable. These amount to the $3.1$\%, $8.3$\%
and $4.6$\% for the \wb, \wv\ and \wi\ samples, respectively.  While the
procedure bias our samples against LBGs that are unresolved by the ACS, it
prevents contamination by galactic stars. In practice, it results in
negligible changes to the measured specific luminosity density. Furthermore,
while the vast majority of galaxies at $z\ll 6$ have \wi-\wi$<1.3$,
photometric errors scatter some of these galaxies into our selection window.
Thus, we have also purged the \wi--band sample from photometric--scatter
interlopers using a statistical procedure. We have used a bright sub--sample
of field galaxies, for which the photometry is very accurate, as a template of
the color distribution of the sample as a whole and estimate the contamination
using photometric errors as function of magnitude derived from the Monte Carlo
simulations in each band (see D03). 

Down to \wz$\le 26.5$, roughly the $50$\%--completeness flux limit for
unresolved sources (see Figure 4 of Giavalisco et al. 2003), the culled 
samples include 1115, 275 and 122 \wb--band, \wv--band and \wi--band dropouts,
respectively. With a survey area of 316 arcmin$^2$, this corresponds to
surface density $\Sigma=3.50\pm 0.10$, $0.87\pm 0.05$ and $0.39\pm 0.03$
galaxies per arcmin$^2$, respectively, where the error bars have been derived
assuming Poisson fluctuations.

\section{THE UV COMOVING SPECIFIC LUMINOSITY DENSITY}

Because no systematic spectroscopic observations of the GOODS are available
yet, we have used Monte Carlo simulations to estimate the redshift
distribution function of the three samples. The technique consists of
generating artificial LBGs distributed over a large redshift range (we used
$2.5\le z\le 8$) with assumed distribution functions of UV luminosity (we used
a flat distribution, see later), SED, morphology and size. We have adjusted
the input distribution functions of SED and size by requiring that the
distribution functions recovered from the simulations using identical
procedures as for the real Lyman--break galaxies match the observation at
$z\sim 4$. In this way, both simulations and observations are subject to
similar incompleteness, photometric errors in flux and color, blending, and
other measurement errors. We have obtained the SED using the unobscured
synthetic spectrum of a continuously star--forming galaxy with age $10^8$ yr,
Salpeter IMF and solar metallicity (Bruzual \& Charlot 2000), and reddened it
with the starburst extinction law (Calzetti 2000) and $E(B-V)$ randomly
extracted from a gaussian distribution with $\mu_{E(B-V)}=0.15$ and
$\sigma_{E(B-V)}=0.15$. For the cosmic opacity, we have used the Madau (1995)
prescription extrapolated to $z=8$. We have used an equal number of $r^{1/4}$
and exponential profiles with random orientation, and size extracted from a
log--normal distribution function, as described by Ferguson et al. 2003, this
volume, Figure 2).

For a given dropout sample, the main output from the simulations is the
probability function $p(M,z,m)$ that a LBG with absolute magnitude $M$ at
redshift $z$ is observed as having apparent magnitude $m$. A commonly used
technique to derive the specific luminosity density of LBGs in this case is
that of the ``effective volume'' (S99), where the spatial volume occupied by
galaxies with apparent magnitude $m$ in the sample is  
$$V_{eff}(m) = \int\int p(M,z,m)\, dM\, {dV(z)\over dz}\, dz.\eqno(2)$$ 
The comoving specific luminosity density contributed by such galaxies is then
estimated as  
$$d{\cal L}(m) = {n(m)\, L(m,\bar z)\over V_{eff}(m)}\, dm,\eqno(3)$$
where ${\bar z}$ is the average redshift of the simulated galaxies that have
been selected into the sample, $n(m)$ are the number counts of real
Lyman--break galaxies observed with magnitude $m$, and $L(m,\bar z)$ is the
specific luminosity of one such galaxy if placed at redshift $\bar z$. 
The specific luminosity density is then ${\cal L} = \int d{\cal L}(m)$.  
The method provides a relatively accurate estimate in the case of the
color--color selected \wb--band and \wv--band dropout samples, regardless of
the assumptions about the UV luminosity distribution function used in the
simulations, because these galaxies are selected based only on their observed
SED, i.e. redshift, with little dependence of their intrinsic luminosity. In
other words, these LBGs have a relatively tight correlation of $m$ and $M$ up
to the the detection limit.

For the one--color selected \wi--band dropouts the $V_{eff}$ method
underestimates the value of ${\cal L}$, because in this case there is no tight
correlation between absolute and apparent magnitude in the sample. Galaxies
with a given apparent magnitude $m$ now have absolute magnitude $M$
distributed in a much larger interval, because whether or not they enter the
sample depends on both their redshift and on their absolute magnitude.
Specifically, the lower redshift bound is set by the color threshold, while
the upper bound depends strongly on galaxy luminosity, as IGM opacity
suppresses the \wz--band flux at higher redshifts. Hence, the effective volume
$V_{eff}(m)$ is overestimated (unless in the simulation one uses the same
intrinsic luminosity function fo the real galaxies), because in the
simulations galaxies with widely different absolute magnitude are equally
represented. In this case we have measured ${\cal L}$ with a different
procedure. We have done a $\chi^2$ minimization to find the intrinsic
luminosity function $\phi(M)$ of the simulated galaxies such that once
inserted into the real images and retrieved their number count
$$n_s(m) = \int\int p(M,z,m)\, \phi(M)\, dM\, dz\eqno(4)$$ 
is best--fit to the number count of the real galaxies. We use a Schechter
function with slope fixed to the value found at $z\sim 3$ (Adelberger \&
Steidel 2000, AS00 hereafter), namely $\alpha=1.6$, and derive the parameter
$M^*$ and $\phi^*$ from the fit. In practice, since our data only reach down
to $L\sim 0.2 L_3^*$ at most, the assumption of a fixed value of $\alpha$ is a
relatively minor one (we verified that values in the range $1.4\le\alpha\le
1.8$ change our results by at most $\sim 12$\%). The specific luminosity
density is then simply ${\cal L} = \int L\, \phi(M)\, dM$, and for each
redshift value this is computed using the corresponding sample's best--fit
luminosity function (a detailed description of the fits and of the analysis of
the result is the subject of a forthcoming paper).

We found the average redshift and standard deviation of the redshift
distribution to be $z_B=3.78$ and ${\cal S}_B=0.34$ for the \wb--dropout
sample, $z_V=4.92$ and ${\cal S}_V=0.33$ for the \wv\ sample, and $z_i=5.74$
and ${\cal S}_i=0.36$ for the \wi\ one. For each sample we have estimated
${\cal L}$ at $1500$ \AA\ (${\cal L}_{1500}$) using K--corrections derived
from a template LBG with the same UV color as the average one in the \wb--band
dropout sample. We have derived ${\cal L}_{1500}$ integrating the luminosity
function down to the ``observed'' limit, i.e. the absolute magnitude that for
each sample corresponds to an apparent magnitude of \wz=26.5, and we have also
estimated the (relatively small) corrections so that all the measures reach
the same absolute magnitude, which we have chosen to be that of the B--band
dropout sample ($0.2\, L_3^*$). Finally, when using the $V_{eff}$ method, we
have also estimated the magnitude--dependent corrections for light losses due
to finite aperture photometry that we have derived from the Monte Carlo
simulations.

The top panel of Figure 1 shows ${\cal L}_{1500}$ for the GOODS three samples
derived using the two methods above, both the ``observed'' values and those
corrected to the same absolute magnitude, compared to the measures at $z\sim
3$ and $\sim 4$ by S99. It can be seen that the two methods provide similar
results for the \wb\ and \wv\ dropout samples, but differ for the \wi--band
dropouts. Note that the luminosity function correction is relatively small for
the \wv--band dropouts, but increases for the \wi--band ones, as the \wz=26.5
limit of the sample corresponds to a brighter intrinsic luminosity at higher
redshift. The bottom panel of the figure shows the star--formation density as
a function of redshift. For the GOODS points we have used the conversion
factor by Madau et al. (1998), namely SFR$=1.4\times 10^{-28}\, {\cal L}$ in
units of \sfr\ Mpc$^{-3}$; the other points are from S99, after conversion to
our world model and correcting for the smaller range of absolute luminosity,
i.e. $>0.2\, L_3^*$ instead of $>0.1\, L_3^*$. Both the observed values and
those corrected for dust obscuration as suggested by AS00 are plotted. 

\section{DISCUSSION AND CONCLUSIONS}

The specific luminosity density of LBGs appears to depend rather weakly on
redshift over the range $2.5\simlt z\simlt 6.5$. Integrating down to $L\sim
0.2L_3^*$, S99 report ${\cal L}_{1500}(z=3)=1.50\pm 0.10$, while from the GOODS
data we find ${\cal L}_{1500}(z=4)=1.63\pm 0.05$, ${\cal L}_{1500}(z=5)=
1.04\pm 0.08$, and ${\cal L}_{1500}(z=6)=1.15^{+0.24}_{-0.19}$ in units of 
$\times 10^{26}$ \cld\ (1--$\sigma$ error bars). However, while the points at
$z\le 5$ are relatively robust, the constraint at $z\sim 6$ is still somewhat
weak, because it critically depends on the measure of the number counts near
the sensitivity limit of the survey (see D03). For example, if we restrict the
sample to \wz--band photometry with $S/N\ge 6.5$ (roughly \wz$<26$) then we
found that the best-fit specific luminosity density drops down to ${\cal
L}_{1500}(z=6)=0.42^{+0.33}_{-0.30}\times 10^{26}$ \cld\
($0.42^{+0.60}_{-0.40}$ 2--$\sigma$). At face value this is still a relatively
mild drop from the $z\sim 3$ value, a factor $\approx 3.5$, smaller than the
factor $\approx 7$ proposed by Stanway et al. (2003), but the error is too
large for the constraint to be meaningful. We note, though, that the $z\sim 5$
point, which is more robust than the $z\sim 6$ one and does not need as big an
extrapolation down the luminosity function, is reasonably consistent with the
mild evolution scenario, lending support to our best estimate at $z\sim 6$.

Thus, the result at $z\sim 6$ depends somewhat critically on the apparently
substantial amount of light contributed by the faint galaxies. To the best of
our knowledge the galaxies with $5\le S/N\le 6.5$ are real $z\sim 6$
candidates. Based on our color selection, the observed dispersion of colors of
faint galaxies, our simulations, and the expected colors of galaxies of
various spectral types, we believe that our statistical corrections to the
contamination is adequate and is not introducing a major systematic error in
our measures. The very good agreement with the measure by S99 at $z\sim 4$,
which is supported by systematic redshift identification of the galaxies, adds
credence to this assertion. The current spectroscopic identifications at
$z\sim 6$ (Bunker et al. 2003; D03; Stanway et al. 2003b) agree well with
our predicted redshift distribution and also support this conclusion. The
samples are still too sparse, however, to attempt a measure of the efficiency
of the selection criteria and, thus, the contamination. Clearly, the measure 
of ${\cal L}_{1500}(z=6)$ clearly needs to be revisited with deeper data. 

In any case, our measure at $z\sim 6$ is in overall good agreement with other
similar measures from {\it HST} and ground--based data (Bouwens et al. 2003;
Lehnert \& Bremer 2003). Stanway et al. (2003a,b) report a factor $\approx 7$
decrease of the specific luminosity density, although this is also very likely
in agreement with our resultof because they limit their measure to bright
galaxies, i.e. $L>L_3^*$. Actually, a direct quantitative comparison is
difficult because of the difference in the sample selection and the pronounced
dependence of ${\cal L}_{1500}$ on the faint counts. These authors also use
the GOODS data, but they base their source detection on single--epoch images
(we use a stack of three epochs), therefore necessarily reaching a shallower
flux level and larger incompleteness\footnote{They also use a slightly
different color selection criterion, namely \wi-\wz$>1.5$. As D03 observe,
this misses one of the galaxies spectroscopically confirmed at $z=5.8$.}. Down
to \wz$<25.6$ they find a total of 14 galaxies over 350 arcmin$^2$ (0.040
arcmin$^{-2}$), while to the same flux level and using their same criteria we
find 35 galaxies over 316 arcmin$^2$ (0.11 arcmin$^{-2}$). Also, note that
they use the $V_{eff}$ method to derive ${\cal L}$, which, as we have detailed
above, underestimates the specific luminosity density for the \wi--band
dropouts.

Our measures are in good quantitative agreement with another deep,
large--area LBG survey at $z\sim 5$ (Iwata et al. 2003) and also with surveys
based on photometric redshifts (Thompson et al. 2001; Thompson 2003; Fontana
et al. 2003; Kashikawa et al. 2003). They differ from some of the photometric
redshift estimates (from the HDF) by Lanzetta et al. (2002), who, after
correcting for bias in the photometry due to the $(1+z)^4$ cosmological
dimming, conclude that the specific star--formation density either remains
constant up to $z\sim 6$, and then increases afterward, or increases
monotonically. While we do explicitly account in our measure corrections for
light losses due to finite aperture photometry (estimated from the Monte Carlo
simulations), we do not find as large corrections as those suggested by
Lanzetta et al (2002). Very likely, much of the difference is that our
corrections are derived by requiring that the galaxy size distribution used as
input in our Monte Carlo simulations is such that the output distribution
matches the observations at $z\sim 4$ (this roughly corresponds to assuming
that the sizes evolve as $\sim H(z)^{-1}$; see Ferguson et al., this volume),
while Lanzetta et al.'s corrections are based on the light distribution of
galaxies at $1\simlt z\simlt 1.5$, i.e. at considerably lower redshifts.

It is not known if the UV specific luminosity density of Lyman--break galaxies
at $z>2.5$ is a fair tracer of the cosmic star formation activity (e.g. AS00).
Regardless, if the dust obscuration properties of these sources are similar to
local starburst galaxies (Meurer et al. 1999; Calzetti 2000), and do not
significantly change over the redshift range $3<z<6.5$, then our result can be
rephrased by saying that the star--formation activity of LBGs decreases very
mildly with increasing redshift, at $z\sim 6$ being $\sim 25$\% lower than it
was around its maximum at $1\simlt z\simlt 3$, several Gyr earlier. Since LBGs
are certainly sites of formation of a considerable amount of the present--day
cosmic stellar mass density (AS00; Giavalisco 2002) it appears that the onset
of substantial cosmic star formation takes place earlier than $z>6$, namely at
less than $\sim 7$\% of the cosmic age. Finally, Figure 1 also shows that
there is an overall good agreement between the dust--corrected data points and
the semi--analytical models by Somerville et al. (2001, blue curve), and
between the uncorrected data points and the predictions based on the observed
evolution of the neutral \HI\ mass density as traced by DLA (which include the
effect of dust obscuration) at least up $z\sim 4$, where these systems are
known with goods statistics (Pei et al. 1999, red curves).

What is the number of hydrogen--ionizing photons expected from our
$i_{775}-$band dropouts at $z\sim 6$? Since the rest-frame UV continuum at
1500 \AA\ is dominated by the same short-lived, massive stars that are
responsible for the emission of photons shortward of the Lyman edge, the
needed conversion factor, about one ionizing photon above 1 ryd for every 5
photons at 1500 \AA, is fairly insensitive to the assumed initial mass
function and is independent of the galaxy history for ages $\gg 10^{7}\,$ yr
(Madau, Haardt, \& Rees 1999; Haiman, Abel, \& Madau 2001). Here we normalize
the number of ionizing photons to the {\it observed} 1500 \AA\ flux
(rest-frame), i.e. we bypass the need for any correction due to dust
extinction. A comoving luminosity density of ${\cal L}_{1500}=10^{26}\,$erg
s$^{-1}$ Hz$^{-1}$ Mpc$^{-3}$ implies then a comoving production rate of
H-ionizing photons $\dot n_{\rm ion}\approx 3\times 10^{51}$ s$^{-1}$
Mpc$^{-3}$, or about 5$\,f_{\rm esc}$ photons per hydrogen atom per $5\times
10^8$ yr escaping into the intergalactic medium. Here $f_{\rm esc}$ is the
average escape fraction of ionizing radiation from the galaxy \HI layers {\it
relative to the escape fraction at 1500 \AA}. Photoionization of intergalactic
hydrogen requires more than one photon above 1 ryd per hydrogen atom, as extra
photons are needed to keep the gas in overdense regions and filaments ionized
against radiative recombinations. If $f_{\rm esc}$ is greater than a few tens
of a percent, then our population of $i_{775}$ dropouts may contribute
significantly to the UV metagalactic flux, and help to reionize the universe
by redshift 6.

\vskip1cm

Support for the GOODS {\it HST} Treasury program was provided by NASA through
grants HST-GO09425.01-A and HST-GO-09583.01 from the Space Telescope Science
Institute, which is operated by the Association of Universities for Research
in Astronomy, under NASA contract NAS5-26555. PM acknowledges support by NASA
through grant NAG5-11513. We thank the referee, Dr. Adriano Fontana, for a
very careful and thoughtful report.

\newpage
\begin{figure}
\figurenum{1}
\epsscale{0.75}
\plotone{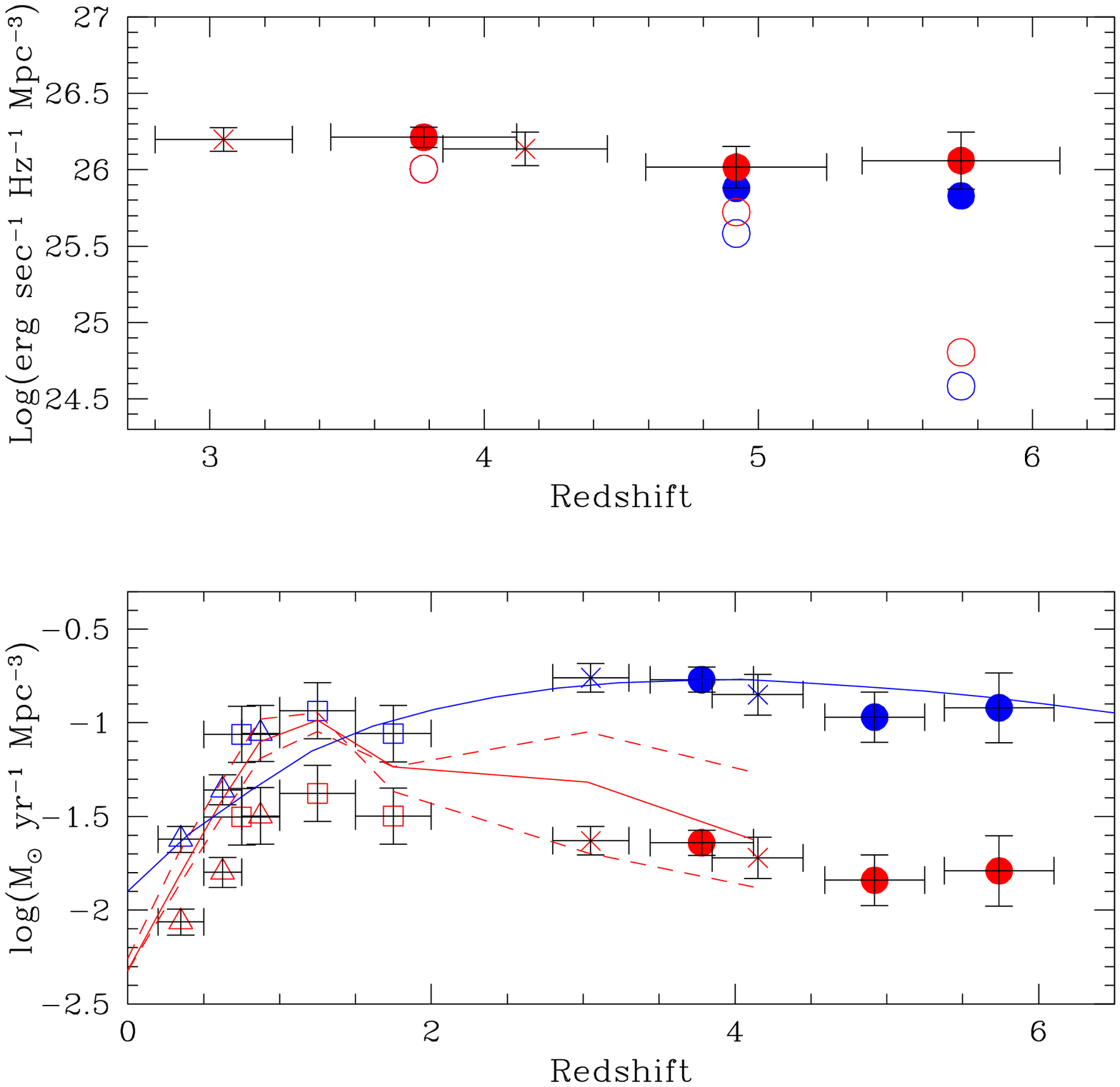}
\caption{{\bf Top.} The specific luminosity density at $\lambda=1500$ \AA\ of
Lyman--break galaxies as a function of redshifts. The circles represent the
\wb--band, \wv--band and \wi--band dropout samples; the crosses are from
Steidel et al. (1999). The hollow circles represent the $V_{eff}$ measures,
the filled circles represent the $\chi^2$ one. The blue symbols are the
measures relative to the observed range of absolute luminosity; the red
symbols include correction down to $L=0.2\times L_3^*$ (see text). The error
bars of the GOODS points are the 68\% confidence interval of the fitting
procedure. {\bf Bottom.} The average star formation density of UV--bright
star--forming galaxies as a function of redshift. The GOODS points have been
obtained from the specific luminosity density using the conversion factor by
Madau et al. (1998). The other points are from Steidel et al. (1999) after
conversion to our world model. The red points are as observed, the blue points
have been corrected for dust obscuration using the procedure proposed by AS00.
The red curve is derived from the evolution of the \HI\ mass density as traced
by DLA absorbers (Pei et at. 1999; the blue curve is from semi--analytical
models (Somerville et al. 2001).}
\end{figure}
\end{document}